%% file: SpiralGalaxyStructure-update.tex
\documentclass[10pt,twocolumn,letterpaper]{article}

\usepackage{cvpr}
\usepackage{times}
\usepackage{epsfig}
\usepackage{graphicx}
\usepackage{amsmath}
\usepackage{amssymb}

\usepackage{nicefrac}

\usepackage[caption=false,lofdepth,lotdepth]{subfig}

\usepackage[pagebackref=true,breaklinks=true,letterpaper=true,colorlinks,bookmarks=false]{hyperref}

\newcommand{\slfrac}[2]{\left.#1\middle/#2\right.}
\newcommand{\resourcePath}{./}

\cvprfinalcopy 

\def\cvprPaperID{1293} 

\ifcvprfinal\fi

\begin{document}

\title{Automated Quantitative Description of Spiral Galaxy Arm-Segment Structure}

\author{Darren R. Davis, Wayne B. Hayes\\
University of California, Irvine\\
Irvine, California 92697-3435 USA\\
{\tt\small drdavis@ics.uci.edu, wayne@ics.uci.edu}\\
\\
April 2013\thanks{This version includes an enhanced Methods section, as well as additional background information and citations that were not included in the original version due to space limitations when submitting to CVPR 2012. As such, additional information covers what was available as of April 2012 (rather than the ArXiv submission date of April 2013), for consistency with the final submission date for CVPR 2012.}
}

\maketitle
\thispagestyle{empty}

\input{SpiralGalaxyStructure-main-text-update.tex}
{\small
\bibliographystyle{ieee}
\bibliography{SpiralStructCVPR2012,CVPRsup}
}

\end{document}

%% file: SpiralGalaxyStructure-main-text-update.tex
\begin{abstract}

We describe a system that builds quantitative structural descriptions of spiral galaxies. This enables translation of sky survey images into data needed to help address fundamental astrophysical questions such as the origin of spiral structure---a phenomenon that has eluded full theoretical description despite 150 years of study (Sellwood 2011). The difficulty of automated measurement is underscored by the fact that, to date, only manually-guided efforts (such as the citizen science project \textnormal{Galaxy Zoo}) have been able to extract structural information about spiral galaxies. An automated approach is needed to eliminate measurement subjectivity and handle the otherwise-overwhelming image quantities (up to billions of images) from near-future surveys. Our approach automatically describes spiral galaxy structure as a set of arcs fit to pixel clusters, precisely characterizing spiral arm segment arrangement while retaining the flexibility needed to accommodate the observed wide variety of spiral galaxy structure. The largest existing quantitative measurements were manually-guided and encompassed fewer than 100 galaxies, while we have already applied our method to nearly 30,000 galaxies. Our output is consistent with previous information, both quantitatively over small existing samples, and qualitatively with human classifications.
   
\end{abstract}

\section{Introduction}
\label{sec:Intro}

In 2003-2004, the Hubble Space Telescope took a one-million-second exposure of \nicefrac{1}{13,000,000}th of the sky. Containing about 10,000 galaxies \cite{Beckwith2006}, the resulting ``Hubble Ultra-Deep Field'' suggests that the entire sky contains upwards of $10^{11}$ galaxies observable from orbit at this exposure depth. The Sloan Digital Sky Survey \cite{York2000} has already imaged almost 1 million galaxies observable from the ground. It will soon be joined by the Vista Hemisphere Survey, Dark Energy Survey, Large Synoptic Survey Telescope, and possibly later by the proposed WFIRST and Euclid space-based missions. With increased sky coverage, increasing sensitivity and resolution, and imaging at different wavelengths, the quantities of galaxy images will continue to increase.

Although spiral galaxies are intrinsically ``fuzzy'' objects subject to highly multifaceted variation (especially in spiral arm structure), they roughly consist of a combination of components: one spheroidal bulge, one flat round disk, an optional linearly-elongated bar, and any number of arc-shaped spiral arms. The first three are each symmetric about a common center; arms tend to resemble a spiral pattern.
\begin{figure}[t]
	\centering
		\includegraphics[scale=0.4]{\resourcePath 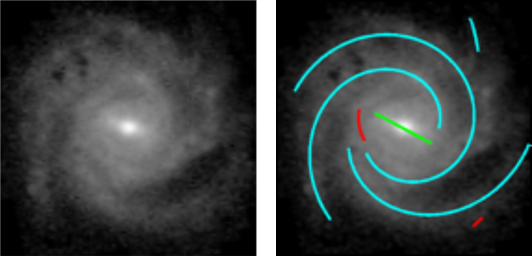}
	\caption{A galaxy from the Sloan Digital Sky Survey with detected counterclockwise (\textit{cyan}) and clockwise (\textit{red}) spiral arm-segments, along with the detected bar (\textit{green}). Our method quantitatively describes spiral galaxy structure as a list of arcs fit to pixel clusters. By parameterizing these ``arm segments'' individually, we quantify arm shape and capture arm-arrangement variations without imposing assumptions from an explicit galaxy-level structure model. Overall structure, and derived measurements (e.g., winding direction, arm tightness, asymmetry), can then be calculated from the arm segments' shape and position parameters.}
	\label{fig:ExampleFit}
\end{figure}

Quantitative structural information about spiral arms would be useful to astronomers in many contexts.
Theory and simulations predict that structure correlates with external variables such the local environment and the cosmological age in which the galaxy resides \cite{Savchenko2011}. Additionally, the pitch angle (tightness) of spiral arms correlates with internal variables such as the rotation velocity of the galaxy \cite{KennicuttJr1981} and the mass of its central black hole \cite{Seigar2008}. Interestingly, no satisfactory explanation exists for what causes spiral arms, due at least in part to a lack of quantitative observational data on their structure \cite{Martinez-Garcia2011,Sellwood2010}. Finally, modern cosmology assumes that the universe as a whole does not rotate, although this assumption has yet to be observationally tested with a sample large enough to provide adequate certainty  \cite{Land2008, Longo2011}.

As discussed in Section \ref{sec:PrevWork}, spiral galaxy structure data currently requires human observers (see, for example, \cite{Lintott2008, Nair2010a}). Automated structure extraction would improve the feasibility of the corresponding studies under increasing data volumes, while enabling new investigations that rely on large-scale, objective quantitative structure measurement rather than subjective, coarse-grained classifications or per-galaxy manual input.

Spiral galaxies exhibit myriad shape variations, making structure description challenging. Spiral arms can exist in any number, and the arm count may not even be well-defined. Spiral arms need not be symmetric, nor conform to any other arrangement. Spiral arms also vary from strongly coherent components that span the full radial extent of the galaxy to small, numerous fragments that only locally exhibit a spiral pattern. These arms also lack consistent spatial relationships, and may not be contiguous with other components.

A mixture model may seem appealing due to categorizations such as arm count, but would face intricate sub-variations. Possible spatial relations between galaxy elements may suggest a pictorial structure model \cite{Felzenszwalb2005}, but spiral galaxies lack a consistent tree structure. Moreover, imposing a shape model risks favoring model structure over galaxy structure, preventing accurate description. Even if a model can accommodate all possibilities (which would require extensive spiral galaxy structure data for verification), it will impose a broad search space. 

Visual grammars (e.g., \cite{Zhu2006}) may have the requisite flexibility to express spiral arm structure, along with the hierarchical relations among components, but the structure-bias and search-space issues would need to be carefully handled. By providing training or comparison data, output from a simpler system could address concerns about structure bias. For example, grammars could describe spiral arm ``forks,'' and output from an independent-arc model could verify that corresponding changes to the arcs are reasonable. Additionally, a simpler system could address search space size by providing initial conditions or training data, and could establish a useful baseline for assessing which model complexity additions are worthwhile. 

In order to perform quantitative analysis of spiral galaxy structure, and to establish a testbed for grammar-based galaxy inference, a system is needed that can describe a wide variety of structural configurations with minimal assumptions, and also provide good per-galaxy initial conditions. No such system previously exists, but our approach can extract such quantitative information and (as discussed further in Section \ref{sec:Conclusion}) facilitate grammar-based inference. Figure \ref{fig:ExampleFit} gives an example image, visualizing the quantitative information we can extract with a list-of-arcs model that avoids imposing arm-arrangement assumptions.



\section{Previous Work}
\label{sec:PrevWork}

The Hubble scheme \cite{Hubble1936} is a historically important subjective classification system that is still in wide use today. Bar presence splits the spiral galaxy portion into two parallel sets of three to four categories arranged linearly according to arm definition, central bulge dominance, and arm tightness. Many automated classifiers can determine Hubble type \cite{Ball2009a, Banerji2010b}, but the categories themselves discard important structural information since they codify subjective visual observation, discretize continuous quantities, discard several types of structure variation and allow conflicts among classification criteria.


As a rough astronomical analogue of part-based models, other methods explain the observed light through a set of luminosity-adding components, often using parameterizations of the S\'{e}rsic profile \cite{Sersic1963} 
\begin{equation}
\label{eq:SersicProfile}
I(R) = I_e \exp \left\{ -k_n \left[ \left( \slfrac{R}{R_e} \right) ^{1/n} - 1 \right] \right\}
\end{equation}
where $R_e$ is the radius containing half of the total light, $I_e$ is the brightness at that radius, $n$ is the S\'{e}rsic index (controlling the brightness concentration), and $k_n$ is an auxiliary function of $n$. This models the decrease in brightness as a function of distance from the galaxy center; it is extended to two dimensions by assuming that brightness contours are concentric generalized ellipses (with additional free parameters). Systems including BUDDA \cite{DeSouza2004}, GIM2D \cite{Simard1998}, and GALFIT \cite{Peng2002a} use this paradigm to model galaxy brightness as a central, spheroidal bulge plus a flat disk. GALFIT and BUDDA can include a stellar bar, and work is in progress to use multiwavelength data \cite{Perret2009,Bamford2011}.

Such models can be found automatically, but cannot describe the non-ellipsoidal shape of spiral arms. GALFIT \cite{Peng2002a} has recently \cite{Peng2010} introduced non-symmetric components where generalized-ellipse contour radius can be modified as a function of polar angle, along with optional spiral rotations of the coordinate system. By accommodating an arbitrary number of these brightness components, each with their own shape parameters, GALFIT can describe a wide range of spiral structure. However, the fitting (a least squares minimization on the $\chi^2$ model-to-image distance) requires a human to carefully specify initial values for all model parameters, along with all the model components and settings. This makes GALFIT best suited for precise refinement of a manually constructed model, but our system's output could ultimately enable automatic initial conditions for GALFIT. 

Human judgment is still required for measuring quantities such as spiral arm count and tightness. A two-dimensional fast Fourier transform can find the maximum Fourier amplitude as a function of mode (arm) count and spiral arm pitch angle (tightness) \cite{BenDavis2012}; the method assumes spiral-arm symmetry and requires human input \cite{Seigar1998,BenDavis2012}. Pitch angle can also be estimated manually \cite{Ma2001}. 

Au \cite{Au2006} computes local image orientations, then fits a symmetric, two-armed, optionally-barred spiral galaxy model. The model works well for galaxies that have two symmetric arms. However, many galaxies have non-symmetric arms, and galaxies vary in arm count. Yosiphon \cite{YosiphonThesis2009} fits a two-armed grammar-based model using the Expectation-Maximization algorithm \cite{Dempster1977}. Ripley and Sutherland \cite{Ripley1990} describe arms as line-segment chains, but the arms must be attached to a bar or core, and their initial positions must be specified manually (despite attempts to do it automatically). Ganalyzer \cite{Shamir2011} looks for spiral structure by finding intensity peaks in an angle-vs-radial distance plot. It is very fast, but is oriented toward producing a continuous measure of spiral/elliptical classification.

Entirely manual classifications also exist. Galaxy Zoo \cite{Lintott2008} coordinates about 250,000 volunteers to classify SDSS \cite{York2000} images by eye over the Web. The initial Galaxy Zoo project has six categories, including three for spiral galaxies: clockwise, anticlockwise or edge-on \cite{Lintott2008}. Galaxy Zoo 2 adds detail including spiral arm count and tightness, bulge dominance, and bar presence. Recently, one professional astronomer painstakingly constructed a detailed structural catalog for 14,034 galaxies \cite{Nair2010a}. In \cite{Longo2011}, 15,158 winding direction classifications were made by eye.

\section{Methods: Automated Structure Extraction}
\label{sec:Methods}

We accept images that give the telescope's flux measurements (as FITS files), or that have been processed for human viewing (e.g., PNG). In the former case, images have a very high dynamic range, so we rescale them using the function 
\begin{equation}
\label{eq:ImgRescale}
r(x) = \nicefrac{\operatorname{asinh}(\frac{x - m}{\beta})}{\operatorname{asinh}(\frac{M - m}{\beta})}, m \le x \le M.
\end{equation}
Values outside the range [m, M] are clipped. Astronomers frequently use the $\operatorname{asinh}$ function because it is approximately linear for smaller brightness values (bringing out faint details) and approximately logarithmic for larger values (preventing bright regions from dominating the image); the parameter $\beta$ controls the transition point \cite{Lupton2004}. Due to intensity scale variations, we use a linear transformation mapping two brightness quantile levels to $m = 0$ and $M = 100$, and then apply $r(x)$. Exact settings for $\beta$ and the brightness quantiles can depend on the image set (but need not be determined for individual images), and are discussed in Section \ref{sec:Results}.

To remove variation from viewing angle and apparent size, we then standardize the image so that the galaxy's detected elliptical outline is circular and inscribed within square image dimensions of consistent size (256x256 for the experiments in Section \ref{sec:Results}, but other resolutions are also effective). The elliptical outline is determined from an iteratively fitted two-dimensional Gaussian. At iteration $k$ and with image intensities $I_{ij}$, we define pixel weights as
\begin{equation}
w_{ij}^{(k)} = \begin{cases}
  I_{ij} & k = 0 \\
  I_{ij} \cdot \mathcal{N}( \left[ \begin{smallmatrix} i \\ j \end{smallmatrix} \right] |\mu^{(k-1)}, \Sigma^{(k-1)}) & \text{otherwise}
\end{cases}
\end{equation}
where $\mathcal{N}(x |\mu, \Sigma)$ is the standard Gaussian density function. Using these weights $w_{ij}^{(k)}$ over all pixel positions, $\mu^{(k)}$ and $\Sigma^{(k)}$ are then the weighted mean and weighted covariance. To avoid collapsing to the core of the galaxy, we stop when the ellipse fit temporarily stabilizes around the disk. With $w^{(k)} = \sum\limits_{ij} w_{ij}^{(k)}$, we define 
\begin{equation}
c^{(k)} = w^{(k)} - 2 \cdot w^{(k-1)} + w^{(k-2)}, k \geq 3.
\end{equation}
We use the fitted Gaussian at iteration $k$ if $k \geq 3$, $c^{(k)} \geq c^{(k-1)}$ and $0 \geq c^{(k)} \geq -\delta$, using an empirically determined value of $\delta = 0.005$. The $10^{-9}$ contour of the Gaussian likelihood then consistently traces the outline of the galaxy's disk,\footnote{Bright foreground stars may disrupt this process, but astronomers already have methods to subtract them, so we do not spend additional computational time on a star/galaxy mixture model.} which can then be mapped to a consistent shape.

Next, we compute pixel-level orientation features using the approach in \cite{Au2006}. Full, half, and quarter-resolution images are convolved with nine orientation-sensitive filters (one-dimensional Laplacian of Gaussian functions extended along nine orientations). Responses are combined into per-pixel strengths and orientations, and merged across resolutions to give multi-scale information. However, this scheme responds only weakly when arm-disk contrast is subtle, so we first apply an unsharp mask. An appropriate scale for the Gaussian blur ($\slfrac{1}{10}$th of the disk-standardized image size) favors light from the arms over light from the disk, since light from the disk varies at lower spatial frequencies. A future possibility is to subtract a fitted bulge and disk (discussed in Section \ref{sec:PrevWork}). This can be done automatically using existing astronomical software, but such software only works on flux-measurement images.

The orientation filters respond both to galactic bars and spiral arms, since both are locally linear. Since we want separate detection of bars and arms, we first determine bar presence using a Hough transform \cite{Duda1972}. 

The Hough transform responds to a bar (if present), but also to other collinear brightness patterns, requiring (1) identification of lines that might correspond to a bar, and (2) determination of whether any of these lines indeed match a bar. To address the first issue, we compute the Hough transform within a circular region. This favors line segments that intersect the galaxy center---a defining property of bars.
Using the centering from the image standardization step and vote weights from the orientation strengths, we compute the Hough transform within concentric circular regions, producing accumulator matrices $A^{(r)}$ for each integer radius $r$. This can be computed as a single Hough transform by changing the pixel vote order. Each circular region is given the score $\nicefrac{\underset{i, j}{\operatorname{max}} A_{ij}^{(r)}}{r}$, with the first and second matrix indexes corresponding to the distance and angle parameters of the Hough transform for lines. The radius penalty $r$ normalizes against larger radii gathering more votes, and prefers the largest radius containing a line extending bidirectionally to opposite ends of the circular boundary. Since bar length is symmetric in both directions from the core, this further enhances preference for bars (at the possible expense of hypothetical asymmetric bars). We then use the maximum-score radius, excluding pathological low-radius regions where the core produces strong linear votes in all directions with a small radius penalty. In particular, we exclude regions where $\nicefrac{\underset{i, j}{\operatorname{max}} A_{ij}^{(r)}}{\underset{j}{\operatorname{min}} ( \underset{i}{\operatorname{max}} A_{ij}^{(r)} ) }$ falls below an empirically determined threshold of 1.5.

For the second issue, we note that the maximum score designates the bar candidate, including its length (calculated from the region radius) and angle (determined from the highest-scoring Hough transform bin). However, since the orientation field is designed for sensitivity to locally linear patterns, it can produce false detections from core/arm alignments. To test whether a linear brightness pattern is visible in the image itself, a second Hough transform is computed from the image intensities, using a circular region with the radius determined from the previous step, and producing accumulator matrix $B$. The image intensities come from the image standardization mentioned above (skipping the circularization step, since this can make the core appear linear). A second bar score is then computed as $\nicefrac{\underset{i}{\operatorname{max}} B_{ic}^{(r)}}{\underset{i}{\operatorname{max}} B_{ik}^{(r)}}$, where the indexes $c$ and $k$ correspond to the angles closest and perpendicular to the candidate bar angle. Finally, the bar candidate is accepted (and used during subsequent arc-finding) if both scores exceed empirically determined thresholds (7 and 2, respectively).

After bar detection, we describe the spiral arms as a set of logarithmic spiral arcs taking the polar-coordinate form 
\begin{equation}
\label{eq:LogSpiral}
\operatorname{lgsp}(\theta) = r_0 \cdot e^{-a \cdot (\theta - \phi)}, \quad \theta_l \leq \theta - \phi \leq \theta_h
\end{equation}
where $r_0$ is the initial radius, $a$ is the pitch angle (the constant angle between the arc and origin-centered circles), $\phi$ rotates the arc about the origin, and $\theta_l$ and $\theta_h$ define the arc's endpoints. Such logarithmic spirals have been used in previous (manually guided) efforts to describe spiral structure (e.g., \cite{Ma2001,Seigar2006}). Arms with non-constant pitch angle can be modeled using multiple spiral arc segments.

It may be tempting to use an arc-based Hough transform for spiral arm detection, but this proves problematic. False detections can arise from coincidental alignments of unrelated regions, including fragmentary spiral arms. Tightly wound arcs also persist for more revolutions before exiting the image range, gathering more Hough transform votes (even when transforming to a space where line detection can be used, since lines must wrap around). Attempted corrections bias the vote by radius or arm width (a tightly wound arc has more wraparounds within a tight, wide arm). Arm-width influences could be addressed by using edge detections, but arm boundary irregularities can substantially increase noise. Additionally, for a fixed pitch angle, there is a range of ($\phi$, $r_0$) combinations describing the same arc, duplicating Hough transform votes across multiple bins. The extent of duplication depends on the arc endpoints, which are not known \textit{a priori}.

We instead combine bottom-up clustering (which isolates regions likely to belong to the same spiral arm and allows independent arc-fitting) with feedback from the logarithmic spiral model (which prevents clusters from containing multiple arms). For the clustering we define the symmetric pixel-to-pixel similarity between 3x3-neighborhood-adjacent pixels as the absolute dot product of their orientation vectors.
Nearby pixels in the same spiral arm tend to have similar orientations, so we can apply single-link hierarchical agglomerative clustering \cite{Sneath1957, Jain1999} to identify groups of pixels that are likely to belong to the same spiral arm. Cluster similarities are the maximum similarity between an inter-cluster pixel pair.
Initially, each pixel constitutes its own cluster. Clusters merge in similarity order until the next similarity pair falls below a fixed threshold (empirically determined as 0.15). This does not require component-count specification, and as we see below, we can incorporate arc-shape information in addition to local features.

For each cluster $C$, we can fit a logarithmic spiral arc via a nonlinear least squares minimization of the error function 
\begin{equation}
\label{eq:LgspFitErr}
E_{C}[\phi, a, r_0] = \slfrac{\displaystyle\sum\limits_{i \in C} v_i \cdot (\rho_i - \operatorname{lgsp}(\theta_i))^2}{\displaystyle\sum\limits_{i \in C} v_i}
\end{equation}
where $v_i$ is the image intensity at pixel $i$ and $\operatorname{lgsp}(\theta)$ is the logarithmic spiral function defined in Equation \ref{eq:LogSpiral}. $(\rho_i, \theta_i)$ are the polar coordinates of pixel $i$, using the center determined during the image standardization step. We can accommodate multiple revolutions during $\theta_i$ calculations by identifying contiguous inner and outer pixel regions. 

When computing the least-squares fit in practice, we do not need to include $\phi$. For any $\phi$ value outside the $\theta$-range of the cluster pixels (or the inner or outer region for a multiple-revolution arc), the optimal pitch angle and fit error will be the same, with the corresponding optimal $r_0$ determined by $\phi$. Furthermore, once $\phi$ is fixed to an arbitrary value within this acceptable range (in our work, the midpoint of $\theta$-values outside the cluster range), the optimal initial radius for a given pitch angle can be derived analytically. The fit then becomes a fast, single-dimensional optimization on the pitch angle. The arc segment bounds can easily be determined as $\theta_l = \underset{i \in C}{\operatorname{min}} (\theta_i - \phi)$ and $\theta_h = \underset{i \in C}{\operatorname{max}} (\theta_i - \phi)$. 

If one builds the underlying clusters purely from orientation field similarities,
the clusters follow spiral arms, but may combine regions best modeled by multiple arcs (e.g., arm forks, bends or straight regions). We can incorporate arc-shape information in a manner similar to the line-shaped cluster identification in \cite{Kamvar2002}, at the reasonable expense of assuming that individual spiral arms locally follow a logarithmic spiral curve. By rejecting merges with a poor combined arc-fit, we reduce sensitivity to the clustering's stopping threshold (since merges are partially determined by arc fit, rather than maximum pixel similarity alone), and reduce cluster fragmentation without merging two arms into the same cluster. One could use recursive partitioning (e.g., Normalized Cuts \cite{Shi2000a}), but cluster refinements can be more informative than combinations of larger clusters. Furthermore, arm-size variations make cluster-size balancing undesirable, and our method corresponds directly to a model where data are generated along logarithmic spiral arcs, for the same reasons as the line model in \cite{Kamvar2002}. We still preserve modeling flexibility; arms that deviate from our logarithmic spiral model (e.g., with a bend where the pitch angle changes) are naturally split into multiple arm segments. 

When two clusters $C_1$ and $C_2$ are about to merge, we fit a logarithmic spiral arc to the combined cluster, and assess the error increase for each constituent cluster. In particular, we use Equation \ref{eq:LgspFitErr} to calculate the merge score
\begin{equation}
\label{eq:MergeScore}
G[C_1, C_2] = \operatorname{max} \left( \frac{E_{C_1}[\gamma^{(m)}]}{E_{C_1}[\gamma^{(1)}]}, \frac{E_{C_2}[\gamma^{(m)}]}{E_{C_2}[\gamma^{(2)}]} \right)
\end{equation}
where $\gamma^{(i)}$ is the parameter vector $(\phi^{(i)}, a^{(i)}, r_0^{(i)})$ obtained from the least-squares fit to $C_i$. For $\gamma^{(m)}$ we perform a least-squares fit to the merged cluster $C_1 \cup C_2$, with the weights $v_i$ in Equation \ref{eq:LgspFitErr} re-scaled so that $\sum_{i \in C_1} v_i = \sum_{j \in C_2} v_j$. If $G[C_1, C_2]$ exceeds an empirically determined threshold of 2.5, the combined fit is unlikely to properly model at least one of the two clusters, so the merge is skipped. To save computation time and to avoid fitting until a reasonable cluster shape can be determined, we only check this condition when both clusters have reached a minimum size, set to $\frac{R}{10}$, where $R$ is the image resolution in the shorter dimension. 

Clusters can also grow within the bar region. To avoid disruption of cluster-to-arm-segment correspondence, we use our earlier determination of bar presence and location. The bar-fitting error $B_C$ is computed as the image intensity-weighted mean of Euclidean distance transform values \cite{Breu1995} from the line segment defined by the bar position and radius, when a bar detection was made earlier (otherwise, $B_C = \infty$). If $B_C < E_C[\gamma]$, we consider the cluster as part of the bar and substitute $B_C$ for $E_C[\gamma]$ in Equation \ref{eq:MergeScore}.

The resulting clusters correspond to pieces of spiral arms (and the bar, if present), but may be over-fragmented. Obscuring elements (such as dust lanes) can make clusters close but non-adjacent, and increasing the pixel-similarity neighborhood size would vastly increase the number of nonzero similarities. Even if a sufficiently large neighborhood size were feasible, brightness gaps often indicate genuine separations between structure components, so gap-crossing is undesirable until larger-scale arc shape can be assessed. The arc-fit criterion may also reject a merge between two clusters that later grow in a way that significantly improves their merge suitability. Merge-check order is defined by cluster similarities,
and it can be prohibitively expensive to continuously re-check all cluster pairs. However, at the end of the main clustering process, the number of clusters is small enough to conduct purely arc (and bar) model-based clustering. In particular, we calculate merge scores $G[C_1, C_2]$ for all clusters with maximum Euclidean distance $\frac{R}{20}$ between the closest inter-cluster pixel pair, and merge these clusters in score order until no pair's merge score remains below the previously mentioned threshold of 2.5.

At the end of this process, we have a set of pixel clusters, each with a corresponding logarithmic-spiral or bar parameterization. This structure description can then be used to calculate quantities such as spiral arm winding direction and tightness. In the next section, we compare such measurements to previous classifications, both by humans and by partially-automated processes.

\section{Results}
\label{sec:Results}


Galaxy Zoo (GZ) \cite{Lintott2008} is one of the largest-scale sources of information about spiral galaxy structure. We use a sample of 29,250 galaxies, imaged from SDSS \cite{York2000}, with sufficient human votes for categories indicating spiral feature visibility in either Galaxy Zoo 1 (GZ1) \cite{Lintott2010} or Galaxy Zoo 2 (GZ2) \cite{Masters2011}. We rescale the FITS-format images according to Equation \ref{eq:ImgRescale}, with brightness quantiles of 0.2 and 0.999, and $\beta = 2$.

\begin{table*}[tb]
\footnotesize
\begin{center}
\begin{tabular}{|l|c c c c c c |c c c c c c |}
\hline
\textbf{Min Discernibility Rate} & 0 & 60 & 80 & 90 & 95 & 100 & 0 & 60 & 80 & 90 & 95 & 100 \\ \hline
\textbf{Require Longest 2 Agree} & N & N & N & N & N & N & Y & Y & Y & Y & Y & Y \\ \hline \hline
\textbf{Inclusion Rate} & 99.3 & 95.1 & 78.9 & 52.2 & 32.0 & 12.4 & 67.0 & 64.7 & 55.7 & 39.0 & 25.1 & 10.0 \\ \hline \hline
\textbf{Majority Vote} & 75.4 & 75.9 & 77.4 & 79.9 & 81.8 & 82.5 & 79.5 & 79.8 & 81.0 & 83.1 & 84.6 & 84.8 \\ \hline
\textbf{Longest Arc Alone} & 84.9 & 85.3 & 86.7 & 89.4 & 91.3 & 92.4 & 95.3 & 95.7 & 96.5 & 97.8 & 98.3 & 98.4 \\ \hline
\textbf{Length-weighted Vote} & 89.4 & 89.9 & 91.3 & 93.5 & 95.0 & 95.7 & 94.9 & 95.3 & 96.1 & 97.5 & 98.2 & 98.4 \\ \hline
\end{tabular}
\end{center}
\caption{Winding-direction agreement with human classifications from Galaxy Zoo 1. Row 1: the minimum proportion of human votes that the dominant winding direction must receive (out of the two known-direction and four other categories). Row 2: whether we demand that our two longest arcs agree in winding direction. Row 3: the proportion of the 29,250 galaxies included under these criteria. Rows 4, 5, and 6: agreement rates between Galaxy Zoo and three methods of determining winding direction from our output.}
\label{tab:ChiralityAgreement}
\end{table*}
In GZ1, volunteers chose one of six categories for each galaxy, including two for winding direction: Z-wise spiral galaxy and S-wise spiral galaxy. Since multiple voters saw each galaxy, we can calculate two measures of human certainty: winding direction agreement (the proportion of maximum to total winding-direction votes) and discernibility (the proportion of maximum winding direction votes to total votes).
For the images used in this work, we find that humans agree when the winding direction is discernible: when at least 50\% of votes fall within one of the winding-direction categories (which happens for nearly all 29,250 galaxies, since images were included only when humans indicated that features were at least vaguely visible), winding direction agreement is already at least 90\% for 98.4\% of the images. Human discernibility varies much more substantially. Both findings match intuition; galaxy images can be faint and fuzzy, but if arms are visible, winding direction is clear because all arms usually rotate in the same direction and winding direction is binary.

We thus assess our agreement with GZ1 as a function of human discernibility, as shown in Table \ref{tab:ChiralityAgreement}. Recall that our output consists of a list of arcs representing arm segments. Each arc has a length and a logarithmic spiral parameterization. Since longer arcs are more meaningful, we also include results for the subset where our two longest arcs agree in winding direction. In less than 1\% of the 29,250 galaxies, our implementation detectably failed to produce a result, generally due to problems finding the outline of the galaxy disk, or a lack of response from the orientation filters. These images are not included in the comparison (except as part of the inclusion rates in Table \ref{tab:ChiralityAgreement}). A simple majority vote across arcs, without requiring that the longest two arcs agree, achieves 75.4\% agreement with humans. Over the same image set, agreement jumps to 89.4\% when using arc length to weight the votes. 

As we require higher rates of human discernibility, agreement with GZ1 increases steadily to 95.7\%. If we demand that our two longest arcs agree in winding direction, then all human agreement rates increase; arc-length weighted voting achieves agreement rates ranging from 94.9\% to 98.4\%, depending on the required level of human discernibility. 

Even though winding direction is one of the simplest measures of spiral structure, inclusion rates indicate that this problem is difficult even for humans. These inclusion rates drop steadily and dramatically as the discernibility cutoff increases, despite nearly all galaxies getting 20 to 80 votes, comparison set selection based on human determinations of feature visibility, and one of the two winding direction categories receiving more than half of the votes in nearly all cases. Nevertheless, we achieve 89.4\% agreement across more than 29,000 galaxies. We can raise this agreement above 95\% when humans are most confident in winding direction, and above 98\% when also demanding agreement between our two longest arcs.

In Table \ref{tab:LongoChirality}, we compare our measurements to the winding direction classifications from \cite{Longo2011}, using galaxies that overlap with our GZ sample, since we did not have direct access to the images used in \cite{Longo2011}. In \cite{Longo2011}, each galaxy was randomly assigned to one of five human scanners, who categorized the galaxy as Left (i.e., left-handed spin), Right, or Uncertain. Scanners were instructed to use the latter category unless winding direction was clear, leaving about 15\% of the original sample in \cite{Longo2011}. Thus, this set is comparable with the higher-discernability subset of the Galaxy Zoo sample. Our agreement rate is also similar; the arc-length-weighted vote achieves 91.6\% agreement when the human scanner was sure of the winding direction. When our two longest arcs agreed (which occurred about 71\% of the time), agreement increases to 96.4\%. In \cite{Longo2011}, each galaxy was seen by only one observer, so we cannot measure human agreement.
\begin{table}[tb]
\footnotesize
\begin{center}
\begin{tabular}{|l|c|c|}
\hline
\textbf{} & \multicolumn{1}{l|}{\textbf{All}} & \multicolumn{1}{l|}{\textbf{Longest Agree}} \\ \hline
\textbf{Majority Vote} & 78.6 & 82.5 \\ \hline
\textbf{Longest Arc} & 87.4 & 96.6 \\ \hline
\textbf{Length-weighted Vote} & 91.6 & 96.4 \\ \hline
\end{tabular}
\caption{Comparisons with winding directions from \protect\cite{Longo2011}, for galaxies included in our Galaxy Zoo sample. The last column uses the subset of galaxies where our two longest arcs agree.}
\label{tab:LongoChirality}
\end{center}
\end{table}

We also compare pitch angle (arm tightness) measurements, starting with data that the Galaxy Zoo 2 team kindly provided to us in advance of its publication. Here, where human classifiers indicated that there was ``any sign of a spiral arm pattern'' \cite{Masters2011}, they were asked whether the spiral arms were tight, medium, or loose. Although examples and illustrations were given for each category,
choices are not completely precise, and the categories are too coarse for a direct correspondence with our measurements. However, we can examine the distribution as a whole.

\begin{figure}[tb]
	\centering
		\includegraphics[scale=0.5]{\resourcePath 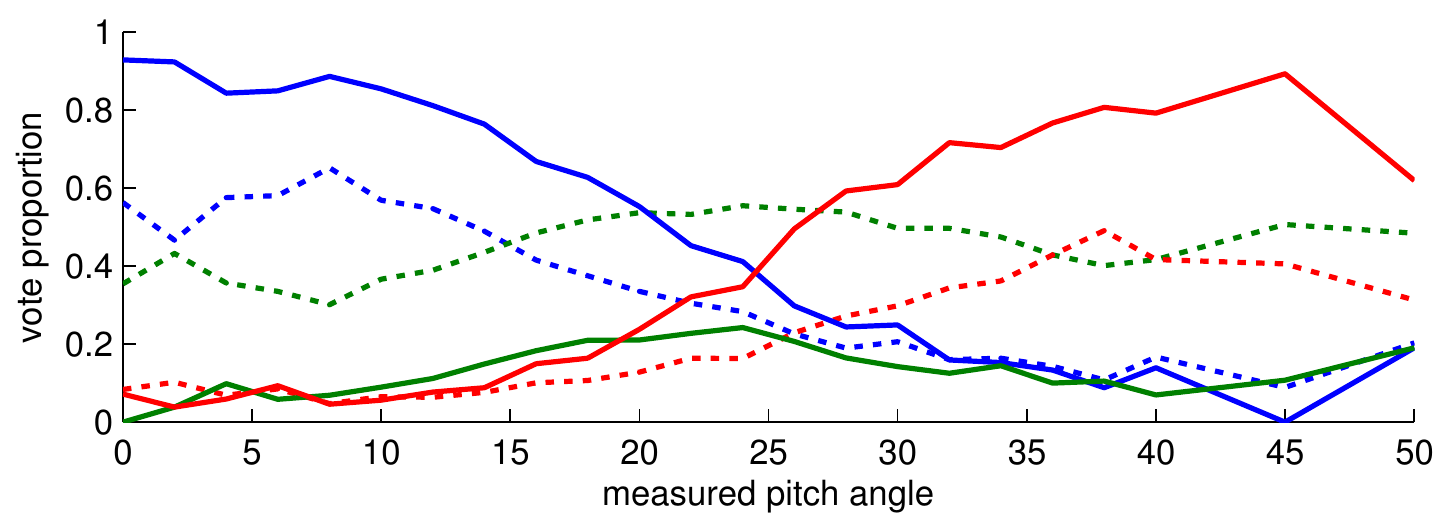}
	\caption{Proportion of galaxies receiving a majority vote for Tight (\textit{blue}), Medium (\textit{green}) or Loose (\textit{red}) as a function of our measured pitch angle, over all images tested from Galaxy Zoo (\textit{dashed}) and for the top human agreement quartile (\textit{solid}). Pitch angles are binned with width 2 degrees from 0 to 40, 5 degrees from 40 to 50, and one bin beyond 50 (due to low sample size).}
	\label{fig:GZ2TightnessComparison}
\end{figure}
Figure \ref{fig:GZ2TightnessComparison} shows the relationship between our measured pitch angle and the proportion of galaxies receiving a majority human vote for Tight, Medium, or Loose. In this and later comparisons, we calculate our pitch angles as an arc-length-weighted sum of pitch angles of individual arcs (the $a$ parameter in Equation \ref{eq:LogSpiral}), using only the arcs that agree with the winding direction indicated by the length-weighted sum of all arcs. As can be seen from the dashed lines in Figure \ref{fig:GZ2TightnessComparison}, galaxies where we measure a low pitch angle usually have majority human votes for Tight, while most of the remaining galaxies in this range had majority votes for Medium. As our measured pitch angle increases, we see progressively fewer galaxies classified as Tight, and more galaxies classified as Loose. Designations as ``Medium'' are pervasive throughout, while ``Loose'' classifications are less frequent than ``Tight'' classifications. This reflects the classification distribution of the image set as a whole.

In the top human agreement quartile (lowest Shannon entropy quartile), the association between our tightness measure and human classifications is even more pronounced, as shown in the solid lines of Figure \ref{fig:GZ2TightnessComparison}. Also, since the Medium majority-votes were far less common with increased human agreement, it seems likely that this choice was frequently used to indicate uncertainty, perhaps due to low galaxy resolution and galaxies with arms of varying tightness. Consequently, it is reasonable that majority-Medium galaxies spread across a wide range of our measured pitch angles. Even if Medium-majority galaxies are disfavored by the entropy measure by having two neighbors (despite most galaxies likely appearing closer to Tight or Loose), such willingness to put many galaxies in any of the three categories would further suggest that much of the spread in classifications stems from human uncertainty. In all, then, we see a clear association between GZ2 tightness classifications and our measurements, with this association strengthening as human agreement increases.

\begin{figure}[tb]
\centering
\subfloat{
\includegraphics[scale=0.35]{\resourcePath 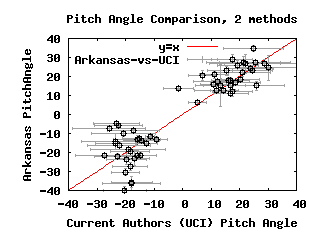}
\label{fig:ArkCmpScatter}}
\subfloat{
\includegraphics[scale=0.35]{\resourcePath 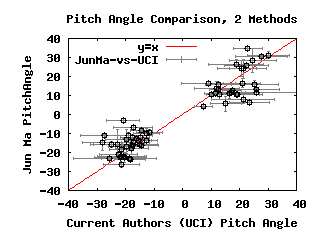}
\label{fig:JunMaCmpScatter}}
\caption{Comparison of our pitch angles with those from a group in Arkansas \protect\cite{Seigar2006,BenDavis2012}, left, and with \protect\cite{Ma2001}, right. In both cases, the vertical axis is the pitch angle measured by the other authors, and the horizontal axis is our measured pitch angle. Our error bars represent a combination of the error inherent in measuring one arm segment and the standard deviation between all arm segments, so these error bars may be overly pessimistic. The Arkansas error bars are described in \protect\cite{BenDavis2012}; error bar widths for \protect\cite{Ma2001} are differences between the two measured arms (when two measurements are available).}
\label{fig:CmpScatter}
\end{figure}
We also compare our measured pitch angles with values calculated using Fourier analysis \cite{Seigar2006,BenDavis2012}. We use R-band FITS images from the Carnegie-Irvine Galaxy Survey \cite{Ho2011}, rescaled using Equation \ref{eq:ImgRescale} with brightness quantiles [0.5, 0.999] and $\beta=1$. This comparison is shown in Figure \ref{fig:ArkCmpScatter}, where a close association can be seen in most cases. Error bars from our method are based on a length-weighted standard deviation of all measured pitch angles agreeing with the dominant winding direction. Since this assumes a Gaussian distribution of arcs' pitch angles (despite the pitch angle distribution clearly having heavier tails when short arcs are included), our error estimations (horizontal bars) are likely to be overly pessimistic. Error measurement calculation from the comparison set is described in \cite{BenDavis2012}. 

In \cite{Ma2001}, pitch angles were determined by manually selecting points along spiral arms, and then fitting one or two logarithmic spiral arcs (depending on visual resolvability). Winding directions (i.e., the signs of the pitch angles) were not listed in \cite{Ma2001}. However, arc fits were displayed for all 60 images, all with unambiguous winding directions, so the second author of this work visually determined winding directions from these arc fits. In our measurements we use images from POSS II \cite{Reid1991}. Since this survey uses photographic plates, with approximately logarithmic rather than linear sensitivity \cite{Lupton2004}, we form a composite of the Infrared, Red, and Blue images, using the procedure described in \cite{Lupton2004}. Comparisons with \cite{Ma2001} are shown in Figure \ref{fig:JunMaCmpScatter}, where we again see a clear association in most cases. Error interval widths for the measurements in \cite{Ma2001} are the differences between measured pitch angles, where two measurements were available. 

\begin{figure}[tb]
\centering
\includegraphics[scale=0.5]{\resourcePath 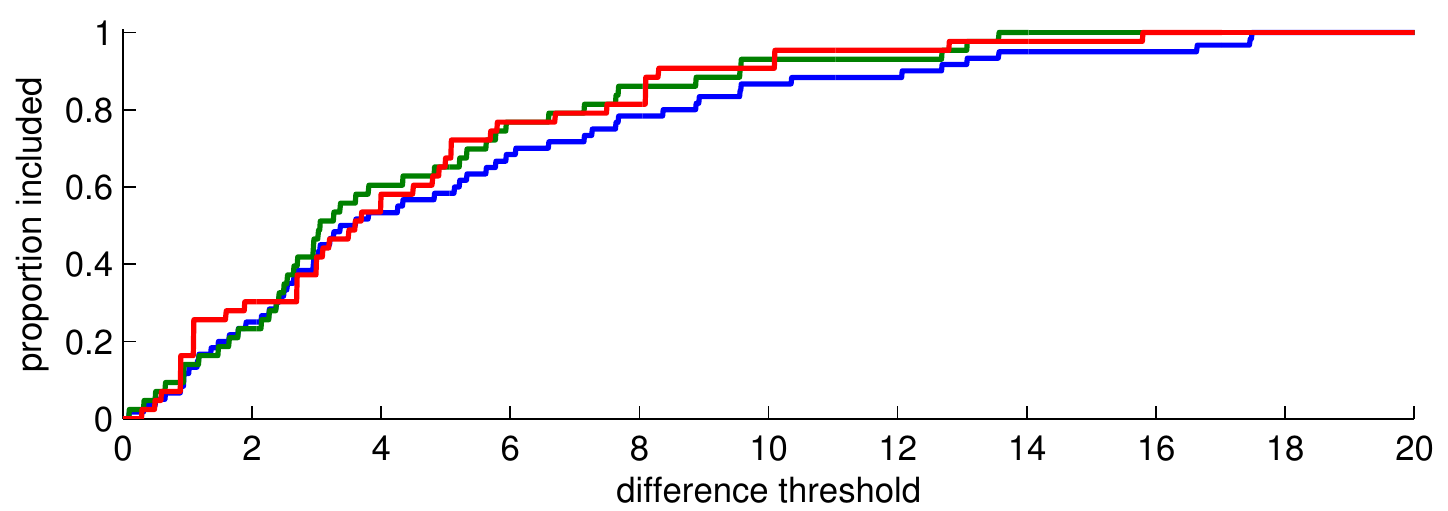}
\caption{Cumulative distribution of within-galaxy pitch angle discrepancies from \protect\cite{Ma2001} (in red) for galaxies where two arcs are measured, and between our measurements and the measurements in \protect\cite{Ma2001} (blue for the full comparison set, and green for the subset where two arc measurements are available from \protect\cite{Ma2001}).}
\label{fig:JunMaDiffVsConsistency}
\end{figure}

In both direct quantitative comparisons of pitch angle, we agree in winding direction, with a strong but not perfect correspondence between pitch angle measurements. Part of the discrepancy can arise because a galaxy's spiral arms can have different pitch angles. We thus compare external discrepancies (differences between our arc-length-weighted mean pitch angles and the mean pitch angles from \cite{Ma2001}) with internal discrepancies (differences between the two measured pitch angles, when available) from \cite{Ma2001}. In Figure \ref{fig:JunMaDiffVsConsistency}, we note the similarity between distributions, especially when comparing only the galaxies where \cite{Ma2001} could derive two arc measurements. Consequently, some between-method differences likely arise from within-galaxy arm variation, especially since we can measure more than one or two arcs.

\section{Discussion and Future Work}
\label{sec:Conclusion}

We have devised a method for fully automatic quantification of spiral galaxy structure, describing individual arm segments even though their count and configuration are not known in advance. This is a challenging task, as indicated by reliance on manual measurement through both crowdsourcing \cite{Lintott2008} and extensive professional effort \cite{Nair2010a,Ma2001,Longo2011}, despite the use of automated methods for other types of structure extraction. Our measurements largely agree with the available manual and semi-manual spiral galaxy structure information, and our work has garnered the interest of multiple astronomical research groups.

At the level of detail we currently extract, most remaining problems arise from galaxy image standardization, and arm-disk contrast. Existing astronomical systems (for galaxy light isolation and disk subtraction) can be used to address both issues. At an average of 90 seconds per core per galaxy on an 2.67 GHz Intel CPU, our prototype Matlab implementation is already fast enough for batch processing of sky-survey data, but we plan to create a C++ implementation for increased speed and deployability. 

Through overall structure distributions and per-galaxy structure configurations, our work can also enable fitting of higher-level models. This would allow extraction of additional detail in a setting with interesting challenges for part-based object recognition (as discussed in Section \ref{sec:Intro}). Thus, we also plan to investigate automatic generation of the initial conditions needed for spiral-fitting in GALFIT \cite{Peng2010}, as well as extensions to visual grammar models (e.g., \cite{Zhu2006}).

Visual grammars would be well-suited to expressing the variable, hierarchical, and recursive structure of spiral galaxies, allowing recovery of finer details and spatial relations between components. Consequently, spiral galaxy structure extraction could be an excellent motivating setting for the development and use of visual grammars. Moreover, our unsupervised, bottom-up part construction can sidestep the need for part annotations by using our clusters as initial information, or by direct extension of our approach. Our method readily enables such an extension, via conversion of our bottom-up clustering into recursive best-split determination (via an algorithm such as the one described in \cite{Wang2006}). Orientation similarities and the arc-fit merge criterion (Equation \ref{eq:MergeScore}) can be readily converted to production rule costs. Subsequent grammar-based approaches could then incorporate astronomical concepts regarding galaxy component relationships and possibly even galaxy formation, using a dynamical grammar model \cite{Mjolsness2006}; see \cite{YosiphonThesis2009} for a 2-armed galaxy model that could be extended and fit to a fuller range of spiral galaxy structure once lower-level information (e.g., arm counts and locations) becomes available through automated methods. With these current and future approaches, we hope to help bridge the gap between large-scale sky surveys and scientiﬁc questions requiring quantitative data about spiral galaxy structure.




\section*{Acknowledgments}
\label{sec:Acknowledgments}

We thank Steven Bamford for helpful insights, image sample selection, and pre-publication access to Galaxy Zoo 2 classifications; Charless Fowlkes, Deva Ramanan, Aaron Barth, and Eric Mjolsness for helpful discussions; and the Arkansas Galaxy Evolution Survey (AGES) Collaboration for pitch angle measurements and discussions. Fellowship, travel, and other support was provided by an ICS Dean's Fellowship at UC Irvine (for DD); iScience at the UC Irvine Donald Bren School of Information and Computer Sciences; the Oxford Centre for Collaborative Applied Mathematics; Steven Bamford and the MegaMorph project; and the AGES Collaboration (through NASA Grant NNX08AW03A).

Comparisons were also made possible due to image data from CGS,\footnote{http://cgs.obs.carnegiescience.edu/CGS/Home.html} as well as from SDSS and POSS II.

\vspace{2mm}

Funding for the SDSS and SDSS-II has been provided by the Alfred P. Sloan Foundation, the Participating Institutions, the National Science Foundation, the U.S. Department of Energy, the National Aeronautics and Space Administration, the Japanese Monbukagakusho, the Max Planck Society, and the Higher Education Funding Council for England. The SDSS Web Site is http://www.sdss.org/.

The SDSS is managed by the Astrophysical Research Consortium for the Participating Institutions. The Participating Institutions are the American Museum of Natural History, Astrophysical Institute Potsdam, University of Basel, University of Cambridge, Case Western Reserve University, University of Chicago, Drexel University, Fermilab, the Institute for Advanced Study, the Japan Participation Group, Johns Hopkins University, the Joint Institute for Nuclear Astrophysics, the Kavli Institute for Particle Astrophysics and Cosmology, the Korean Scientist Group, the Chinese Academy of Sciences (LAMOST), Los Alamos National Laboratory, the Max-Planck-Institute for Astronomy (MPIA), the Max-Planck-Institute for Astrophysics (MPA), New Mexico State University, Ohio State University, University of Pittsburgh, University of Portsmouth, Princeton University, the United States Naval Observatory, and the University of Washington.

\vspace{2mm}

The Digitized Sky Surveys were produced at the Space Telescope Science Institute under U.S. Government grant NAG W-2166. The images of these surveys are based on photographic data obtained using the Oschin Schmidt Telescope on Palomar Mountain and the UK Schmidt Telescope. The plates were processed into the present compressed digital form with the permission of these institutions.

The National Geographic Society - Palomar Observatory Sky Atlas (POSS-I) was made by the California Institute of Technology with grants from the National Geographic Society.

The Second Palomar Observatory Sky Survey (POSS-II) was made by the California Institute of Technology with funds from the National Science Foundation, the National Geographic Society, the Sloan Foundation, the Samuel Oschin Foundation, and the Eastman Kodak Corporation.

The Oschin Schmidt Telescope is operated by the California Institute of Technology and Palomar Observatory.

The UK Schmidt Telescope was operated by the Royal Observatory Edinburgh, with funding from the UK Science and Engineering Research Council (later the UK Particle Physics and Astronomy Research Council), until 1988 June, and thereafter by the Anglo-Australian Observatory. The blue plates of the southern Sky Atlas and its Equatorial Extension (together known as the SERC-J), as well as the Equatorial Red (ER), and the Second Epoch [red] Survey (SES) were all taken with the UK Schmidt.

All data are subject to the copyright given in the copyright summary (http://stdatu.stsci.edu/dss/copyright.html). Copyright information specific to individual plates is provided in the downloaded FITS headers.

Supplemental funding for sky-survey work at the ST ScI is provided by the European Southern Observatory.